\documentclass[prl,aps,twocolumn,showpacs]{revtex4}
\usepackage{epsfig}
\usepackage{graphicx}
\usepackage{textcomp}
\usepackage{amsmath}

\begin{document}
\title{Emergence of stability in a stochastically driven pendulum: beyond the Kapitsa effect}

\author{Yuval B. Simons and Baruch Meerson}

\affiliation{Racah Institute of Physics, Hebrew University of
Jerusalem, Jerusalem 91904, Israel}


\begin{abstract}
We consider a prototypical nonlinear system which can be stabilized by multiplicative noise:
an underdamped non-linear pendulum with a stochastically vibrating pivot. A numerical solution of the pertinent Fokker-Planck equation shows that the upper equilibrium point of the pendulum can become stable even when the noise is white, and the ``Kapitsa pendulum" effect is not at work. The stabilization occurs in a strong-noise regime where WKB approximation does not hold.
\end{abstract}

\maketitle

\begin{figure}[ht]
\includegraphics[width=2.4in,clip=]{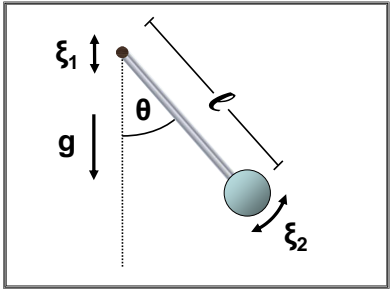}
\caption{ (color online). Schematic of the stochastically driven simple gravity pendulum.
\label{drawing}}
\end{figure}

It has been known for a long time that multiplicative noise can enhance stability of nonlinear systems.
Examples are numerous indeed and culminate at noise-induced phase transitions far
from equilibrium \cite{NIPT}. This paper deals with a noise-induced stabilization of
oscillating systems. As a prototypical example we consider
an underdampled nonlinear pendulum with a stochastically vibrating pivot. The stochastic
driving introduces  both multiplicative and additive noise, see
Fig. \ref{drawing}. Our numerical simulations clearly show that the multiplicative noise can stabilize the otherwise unstable
upper equilibrium point of the pendulum. The mechanism for this stabilization is markedly different from, and more subtle than, the ``Kapitsa pendulum" mechanism. The Kapitsa pendulum involves a (deterministic) monochromatic parametric driving of the pendulum  at a frequency that is much higher than the natural frequency of the pendulum \cite{kap}. Here the upper equilibrium point becomes stable if the driving acceleration is higher than a critical value depending on the pendulum length and the gravity acceleration. In the Kapitsa pendulum problem  the change of stability of the upper equilibrium point comes from a change in the effective potential of the pendulum \cite{kap}.

Extensions of the Kapitsa pendulum effect to multiplicative stochastic driving have also been considered \cite{paper1,paper2}, see Ref. \cite{Ibrahim} for a review. In these extensions the noise spectrum is strongly peaked at a single frequency which is much higher than the natural frequency of the pendulum. The presence of a \textit{high-frequency} noise of a sufficient strength modifies the effective potential which can stabilize the upper equilibrium point. Theory-wise, this setting introduces a time-scale separation which permits a perturbative treatment \cite{Ibrahim}.  In this work we consider a model stochastic driving with a flat spectrum: a white noise. Here all frequencies from $0$ to $\infty$ are equally present, and there is no time-scale separation. We will show that, for such a noise, the upper position of the pendulum can also become stable. However, the stabilization cannot be traced to a change in the effective potential of the pendulum.

A stochastically driven simple gravity pendulum can be described by a Langevin equation:
\begin{align}
&\dot{\theta}=\Omega\,,\label{dottheta}\\
&\dot{\Omega}=-\omega_0^2\sin{\theta}-
2\gamma\Omega+\frac{1}{l}\sin{\theta}
\sqrt{\mu}\,\xi_1(t)+\sqrt{\alpha}\,\xi_2(t)\,,\label{dotOmega}
\end{align}
where $\theta$ is the deviation angle of the pendulum, see Fig.  \ref{drawing}, $\Omega$ is the angular velocity, $\omega_0=\sqrt{g /l}$ is the harmonic frequency of the
pendulum, $\gamma$ is the damping factor, $l$ is the
the pendulum length, $g$ is the gravity acceleration, and $\mu$ and $\alpha$ are the magnitudes of the
multiplicative and additive noises $\xi_1$ and $\xi_2$, respectively. The noises are assumed to be guassian, white with zero mean and mutually uncorrelated:
\begin{equation}
\langle\xi_i(t)\rangle=0\,, \;
\langle\xi_i(t)\xi_j(t^{\prime})\rangle=2\delta(t-t')\delta_{i,j}\,, \; i,j=1,2\,.
\end{equation}
The Langevin equations (\ref{dottheta}) and (\ref{dotOmega}) are equivalent (see, e.g. Ref. \cite{gardiner}) to the following Fokker-Planck equation
for the probability distribution $W(\theta,\Omega,t)$:
\begin{align}
W_t&=-\Omega W_\theta+\omega_0^2\sin{\theta}\,W_\Omega+2\gamma\frac{\partial}{\partial\Omega}(\Omega W)
\nonumber
\\&+\left(\alpha+\frac{\mu}{l^2}\sin^2{\theta}\right)\,W_{\Omega\Omega}\,,
\end{align}
where the indices $\theta$, $\Omega$ and $t$ denote the corresponding partial derivatives of $W(\theta,\Omega,t)$. As the noises are $\Omega$-independent, there is no difference between the Ito and Stratonovich interpretations. Introducing the dimensionless variables $\tilde{t}= \omega_0  t$,
$ \tilde{\Omega} =\Omega/\omega_0 $ and
$\tilde{W}(t,\theta,\Omega) = \omega_0 \, W(t,\theta,\Omega)$, we can rewrite
the Fokker-Planck equation in a dimensionless form:
\begin{align}
W_t&=-\Omega W_\theta+\sin{\theta}W_\Omega+2\Gamma\frac{\partial}{\partial\Omega}(\Omega W)\nonumber
\\&+\left(\varepsilon+\delta\sin^2{\theta}\right)\,W_{\Omega\Omega}\,, \label{1}
\end{align}
where $\Gamma=\gamma / \omega_0$, $\varepsilon=\alpha / \omega^3_0$ and
$\delta=\mu / (l^2\omega^3_0)$ are the rescaled parameters of
the system, and the tildes are omitted.

\begin{figure}[ht]
\includegraphics[width=3.3in,clip=]{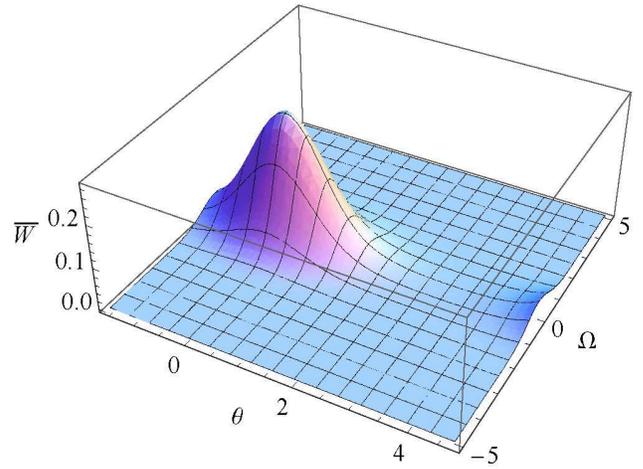}
\caption{(color online). A steady-state probability distribution with no multiplicative noise. The parameters are
$\varepsilon=0.1$, $\Gamma=0.1$ and $\delta=0$. This solution evolved from the initial condition
$W=(2\pi\sqrt{\pi})^{-1}\exp(-\Omega^2)$ after 60 normalized time
units. It coincides with the analytical solution \eqref{chan}.}
\label{figzerodist}
\end{figure}

We assume that, after a transient, the stochastic system approaches a smooth steady state for which $W(\theta, \Omega, t)$ is independent of time: $W(\theta, \Omega, t\to \infty)=\bar{W}(\theta, \Omega)$.
The steady-state probability distribution $\bar{W}(\theta, \Omega)$ obeys the equation
\begin{align}
&-\Omega \bar{W}_\theta+\sin{\theta}\bar{W}_\Omega+
2\Gamma\frac{\partial}{\partial\Omega}(\Omega \bar{W})\nonumber \\
&+\left(\varepsilon+\delta\sin^2{\theta}\right)\,\bar{W}_{\Omega\Omega}=0\,. \label{1steady}
\end{align}
We classify a point ($\theta, \Omega)$ as a stable point of the system if it is a local
maximum of the stationary probability distribution $\bar{W}(\theta, \Omega)$. The necessary and
sufficient conditions for a function of two variables $f(x,y)$ to
have a local maximum at $(x_0,y_0)$ are (see, \textit{e.g.} Ref. \cite{fick}):
\begin{align}
&f_x(x_0,y_0)=f_y(x_0,y_0)=0\,, \label{cond1}\\
&f_{xx}(x_0,y_0)<0 \text{ or } f_{yy}(x_0,y_0)<0\,,\label{cond2}\\
&f_{xx}(x_0,y_0)\,f_{yy}(x_0,y_0)-f_{xy}^2(x_0,y_0)>0\,,\label{cond3}
\end{align}
where the indices $x$ and $y$ denote partial derivatives. Let us examine the
stability properties of the upper equilibrium point ($\theta=\pi,\Omega=0$) of the driven pendulum.
Equation~(\ref{1steady}) is invariant under the transformation
$\theta \to 2\pi-\theta, \Omega \to -\Omega$, that is under reflection of the axes  $\theta$ and $\Omega$ around the point ($\pi,0$).
Its solution $\bar{W}(\theta, \Omega)$
must obey the same symmetry.  Therefore,
the first derivatives $\bar{W}_{\theta}$ and $\bar{W}_{\Omega}$ must vanish at ($\pi,0$), and so Eqs.~(\ref{cond1}) are satisfied there. This immediately follows
$$\bar{W}_{\Omega\Omega}(\pi,0)=-\frac{2\Gamma}{\varepsilon}\bar{W}(\pi,0)< 0,$$
so Eq.~\eqref{cond2} is also satisfied at ($\pi$,0). As a result, the necessary and sufficient condition for ($\pi,0$) to be a stable point is given by Eq.~(\ref{cond3}):
\begin{equation}
\Delta\equiv \bar{W}_{\Omega\Omega}(\pi,0)\bar{W}_{\theta\theta}(\pi,0)-\bar{W}^2_{\theta\Omega}(\pi,0)>0.
\label{D}
\end{equation}
For $\delta=0$ (only additive noise), the steady-state equation (\ref{1steady}) is soluble analytically, see Ref. \cite{chan}:
\begin{equation}
\bar{W}(\theta,\Omega)=\frac{\Gamma^{1/2}}{2\pi^{3/2} \varepsilon^{1/2} I_0(2\Gamma /\varepsilon )} \exp{\left[-\frac{\Gamma}{\varepsilon} \left(\Omega^2-2\cos{\theta}\right)\right]} \,,
\label{chan}
\end{equation}
where $I_0(\dots)$ is the modified Bessel function. In this case the point ($\pi,0$) is unstable, as
the stability parameter $\Delta$, defined in Eq.~\eqref{D},
is negative:
\begin{equation}\label{chan1}
    \Delta_0=-\frac{\Gamma^3}{\pi^3 \varepsilon^3
I_0^2(2 \Gamma /\varepsilon )}
\exp{\left(-\frac{4\Gamma}{\varepsilon}\right)}<0\,.
\end{equation}
The probability distribution (\ref{chan}) is depicted in Fig.~\ref{figzerodist}.

What happens when $\delta>0$? First, we show that there is no stabilization in the weak-noise limit, $\delta \sim \varepsilon \ll 1,\Gamma$. In this limit one can
use a dissipative variant of WKB approximation, see \textit{e.g.} Ref. \cite{gardiner}, and make the ansatz
$$\bar{W}(\theta,\Omega)=A(\theta,\Omega)\exp{\left[-\frac{S(\theta,\Omega)}{\varepsilon}\right]}\,,$$
assuming that the pre-factor $A$ varies on scales much larger than $1/\varepsilon$. In the leading order in $1/\varepsilon$ one obtains
\begin{equation}
\Omega S_\theta-\sin{\theta}S_\Omega-
2\Gamma\Omega S_\Omega+\left(1+\frac{\delta}{\varepsilon}\sin^2{\theta}\right)\,S^2_{\Omega} =0\,.
\label{HJ}
\end{equation}
This first-order PDE for the action $S(\theta,\Omega)$ has the form of a stationary Hamilton-Jacobi
equation  (see \textit{e.g.} Ref. \cite{gold}) with the Hamiltonian $H(\theta,\Omega,p_1,p_2)$ of the form
\begin{equation}
H=\Omega p_1-\sin{\theta}\,p_2-2\Gamma\Omega p_2+
\left(1+\frac{\delta}{\varepsilon}\sin^2{\theta}\right)\,p^2_2\,.
\label{H}
\end{equation}
Here $p_1=S_\theta$ and $p_2=S_\Omega$ are the canonical momenta conjugated to the coordinates  $\theta$ and $\Omega$, respectively. As the Hamilton-Jacobi equation (\ref{HJ}) is stationary,
we are interested in the zero-energy dynamics $H=0$. The Hamilton's equations
\begin{align}
&\dot{\theta}=\Omega\,,\;\;\dot{\Omega}=
-\sin{\theta}-2\Gamma\Omega+2\left(1+\frac{\delta}{\varepsilon}\sin^2{\theta}\right)\,p_2,\\
&\dot{p_1}=\cos{\theta}\,p_2-\frac{\delta}{\varepsilon}\sin{2\theta}\,p^2_2\,,\;\;\;\;\;\;
\dot{p_2}=-p_1+2\Gamma\,p_2,
\end{align}
have two zero-energy fixed points: $a_1=(0,0,0,0)$ and  $a_2=(\pi,0,0,0)$, corresponding
to the lower and upper equilibrium points of the pendulum, respectively. The linear stability of the upper equilibrium point $a_2$ is determined by the quadratic approximation to the Hamiltonian around $a_2$,
\begin{equation}
H(\theta,\Omega,p_1,p_2) \simeq \Omega p_1+\theta p_2-2\Gamma\Omega p_2+p^2_2\,.
\label{Hlin}
\end{equation}
This quadratic approximation is noiseless,  as the noise term in Eq.~(\ref{H}),
$(\delta/\varepsilon)\sin^2{\theta}\,p^2_2$, is bi-quadratic in small deviations from the fixed point. Therefore, a weak multiplicative noise does not generate any correction to the potential of the pendulum, in agreement with an early observation  \cite{paper1}, and cannot change the stability properties of the system.

However, for a strong multiplicative noise numerical solutions of the time-dependent Fokker-Planck
equation \eqref{1} do show emergence of stability of the upper equilibrium point.
We obtained these numerical results
with a ``Mathematica" PDE solver. The numerical domain was $0<\theta<2\pi$ and $-\Omega_{max}<\Omega<\Omega_{max}$
with $\Omega_{max}$ chosen, separately for each set of parameters, sufficiently large. Periodic boundary conditions in $\theta$ were used. As to the boundary conditions in $\Omega$,
we checked that, at sufficiently large $\Omega_{max}$, the steady-state solution remained the same up to $1$ per cent,  whether we imposed
periodic or zero-$W$ conditions at the $\Omega$-boundary. Larger $\Omega_{max}$ needed to be taken when $\Gamma$ became smaller than $\varepsilon$
and when $\delta$ became large compared to
$\Gamma$ and $\varepsilon$. The values of $\Omega_{max}$ that we used ranged
from $\Omega_{max}=2$ for $\Gamma=0.5$, $\varepsilon=0.05$ and
$\delta\simeq0.15$ to $\Omega_{max}=15$ for $\Gamma=0.025$,
$\varepsilon=0.5$ and $\delta\simeq0.43$. After verifying that
$\Omega_{max}$ is sufficiently large, we used the zero-$W$
boundary conditions in $\Omega$, as this choice reduced the computational time.  For each set of parameters $\delta,\Gamma$ and $\varepsilon$ we used the analytical solution \eqref{chan} for $\delta=0$  as the initial condition.

We ran the solution until $t=t_{max}$ such that
the value of $\Delta$, evaluated at $t=t_{max}$, was within 1 per cent from its value at
$t=t_{max}/2$. The larger the parameter $\Gamma$,
the smaller $t_{max}$ was needed. The values of $t_{max}$ that we used
ranged from $40$ for $\Gamma=1$, $\varepsilon=0.3$ and
$\delta=0.6$ to $140$ for
$\Gamma=0.025$, $\varepsilon=0.05$ and $\delta=0.15$.

At small $\Gamma/\varepsilon$ or large $\delta$ the steady-state probability distribution broadens which demands a larger $\Omega_{max}$, a longer computation time and more computer memory. At large
$\Gamma/\varepsilon$ or small $\delta$ the distribution becomes too narrow to numerically resolve it with confidence. We verified that, starting from an arbitrary initial condition, the
probability function always converges to the same steady-state solution. We also
checked that for $\delta=0$ the steady-state solution coincides, with a high accuracy, with the analytical solution~\eqref{chan}.

\begin{figure}[ht]
\includegraphics[width=3.3in,clip=]{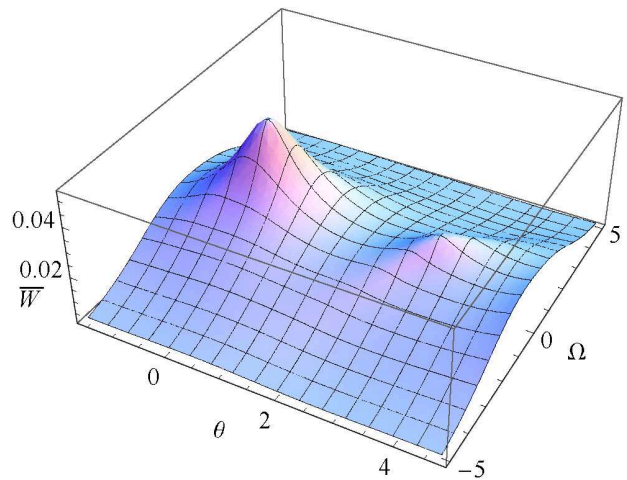}
\caption{(color online). Multiplicative noise stabilizes the upper equilibrium point of the pendulum. Shown is the steady-state probability distribution $\bar{W}(\theta,\Omega)$ which exhibits
a local maximum at the upper equilibrium point ($\theta=\pi, \Omega=0$). The parameters are
$\varepsilon=0.1$, $\Gamma=0.1$ and $\delta=2$. The steady-state  probability distribution was found
by solving numerically the  time-dependent Fokker-Planck equation (\ref{1}). It evolved from the initial condition
\eqref{chan} after $60$ dimensionless time units.}
\label{figstabledist}
\end{figure}

Figure \ref{figstabledist} gives a typical numerical example of a steady-state probability distribution having a distinct local maximum at ($\pi,0$), in addition to the
expected (higher) peak at ($0,0$). Also noticeable is a significant broadening of the probability distribution, compared with the case of only additive noise, see Fig.~\ref{figzerodist}.

\begin{figure}[ht]
\includegraphics[width=3.0in,clip=]{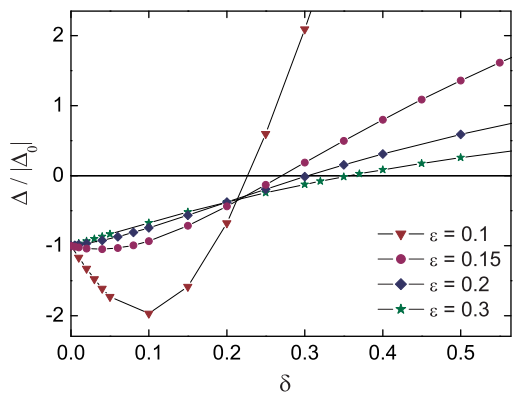}
\caption{(color online). The stability parameter $\Delta$, normalized to $|\Delta_0|$ from Eq.~(\ref{chan1}),  is plotted versus the
rescaled magnitude of the multiplicative noise $\delta$ for different values of $\varepsilon$ at $\Gamma=0.1$.
The lines are only given for guiding the eye. The crossing point of each plot with the line $\Delta=0$ yields $\delta_c$.} \label{figdd}
\end{figure}

To determine the stability properties of the upper equilibrium point ($\pi,0$), we plotted the stability parameter $\Delta$, defined in
Eq.~\eqref{D}, versus the rescaled magnitude of the multiplicative noise $\delta$, for different values of $\Gamma$ and
$\varepsilon$. Several examples of such plots are shown in  Fig.~\ref{figdd}. Using interpolation, we found the critical
values $\delta_c=\delta_c(\Gamma,\varepsilon)$ of $\delta$,  at which $\Delta=0$.

\begin{figure}[ht]
\includegraphics[width=3.0in,clip=]{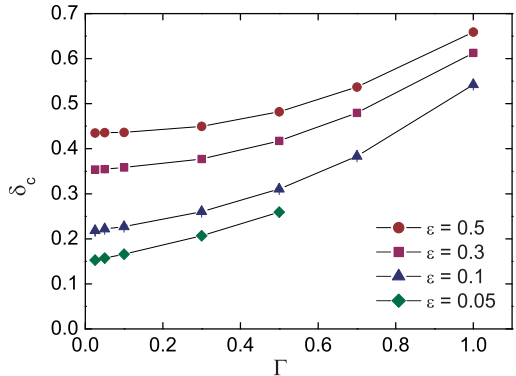}
\caption{(color online). The critical rescaled magnitude of the
multiplicative noise $\delta_c$ vs. the rescaled damping factor
$\Gamma$ for different values of $\varepsilon$.
The lines are only given for guiding the eye.} \label{figgam}
\end{figure}
\begin{figure}[ht]
\includegraphics[width=3.0in,clip=]{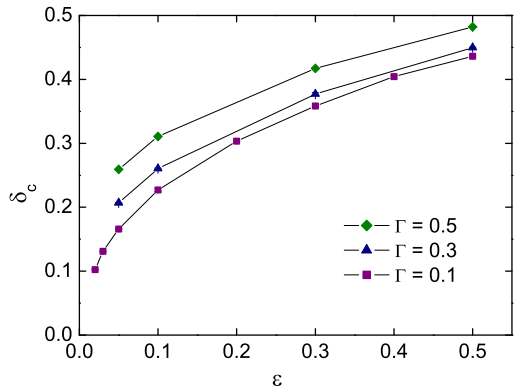}
\caption{(color online).  The critical rescaled magnitude of
the multiplicative noise $\delta_c$ vs. the rescaled magnitude of the
additive noise $\varepsilon$ for different values of $\Gamma$.
The lines are only given for guiding the eye.} \label{figeps}
\end{figure}
\begin{figure}[ht]
\includegraphics[width=3.0in,clip=]{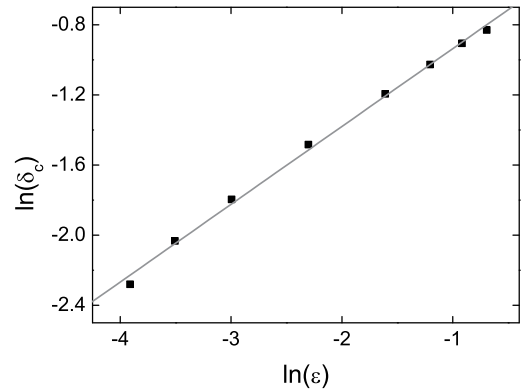}
\caption{The log-log plot of $\delta_c$ vs. $\varepsilon$ for $\Gamma=0.1$. A linear fit gives a
slope of $0.44\pm0.01$ with the coefficient of determination
$R^2=0.99726$.} \label{figln}
\end{figure}

Figure~\ref{figgam} shows the plot of $\delta_c$ versus $\Gamma$ at different
$\varepsilon$.  One can
see that $\delta_c$ slowly decreases with $\Gamma$ and approaches a finite value at
$\Gamma \to 0$. The plot of $\delta_c$ versus $\varepsilon$ at different $\Gamma$, presented in
in Fig.~\ref{figeps}, shows a much stronger, square-root-like dependence at small $\varepsilon$.
Our data for different $\Gamma$ do not contradict  a power-law behavior
$\delta_c \sim \varepsilon^{\alpha}$  with $\alpha\simeq 0.44$, as shown in Fig.~\ref{figln} for $\Gamma=0.1$. When $\varepsilon$ approaches $0$,  $\delta_c$ also goes to zero, and
one would expect to always see stability in this case. As $\varepsilon \to 0$, however, the
probability distribution develops singularities both at
($0,0$), and at ($\pi$,0), but the peak at ($0,0$) becomes much higher than that at ($\pi,0$). A local maximum at ($\pi,0$) is physically insignificant in this case. Only with a sufficiently large
additive noise the probability distribution becomes sufficiently broad to allow a
reasonable probability for the pendulum to be at and around ($\pi$,0).

In summary, we have investigated the stabilization of a prototypical nonlinear oscillating system by a multiplicative white noise.  The stabilization is clearly observed in the numerical solution of the Fokker-Planck equation and requires a super-critical noise magnitude. The stabilization cannot be traced to a change in the effective potential of the system
and is not predicted by a WKB analysis which assumes a weak noise.

We are grateful to Pavel V. Sasorov for useful discussions. This work was supported by the Israel Science
Foundation (Grant No. 408/08).


\begin{thebibliography}{99}
\bibitem{NIPT} W. Horsthemke and R. Lefever, \textit{Noise-Induced Transitions.} (Springer-Verlag, Berlin, 1984).
\bibitem{kap} L.D. Landau and E.M. Lifshitz, \textit{Mechanics} (Pergamon, Oxford, 1960).
\bibitem{paper1} V.E. Shapiro and V.M. Loginov, Phys. Lett. \textbf{71}A, 287 (1979).
\bibitem{paper2} P.S. Landa and A.A. Zaikin, Phys. Rev. E \textbf{54}, 3535 (1996).
\bibitem{Ibrahim} R.A. Ibrahim, J. Vibration and Control \textbf{12}, 1093 (2006).
\bibitem{gardiner} C.W. Gardiner, \textit{Handbook of Stochastic Methods} (Springer, Berlin, 2004).
\bibitem{fick} G.M. Fikhtengol'ts, \textit{The Fundamentals of Mathematical Analysis: Vol. I} (Pergamon, Oxford, 1965).
\bibitem{chan} S. Chandrasekhar, Rev. Mod. Phys. \textbf{15}, 1 (1943).
\bibitem{gold} H. Goldstein, \textit{Classical Mechanics} (Addison-Wesley, 1980).


\end{thebibliography}
\end{document}